\documentclass[epj]{svjour}
\usepackage{latexsym}
\usepackage{graphics}

\def\esym {$E_{\rm sym}(\rho)$}
\def\rpi {$\pi^-/\pi^+$}
\def\rki {$K^+/K^0$}

\def\agev{GeV/u}
\def\amev{MeV/u}
\def\caca{$^{48}$Ca+$^{48}$Ca}
\def\snsn{$^{124}$Sn+$^{124}$Sn}
\def\auau{$^{197}$Au+$^{197}$Au}

\usepackage{amssymb}
\usepackage{amsmath,bm}
\usepackage{graphicx}
\usepackage[normalem]{ulem}
\usepackage[dvips]{color}
\renewcommand\sout{\bgroup \color{red} \ULdepth=-.5ex \ULset}

\begin{document}
\title{Probing nuclear symmetry energy at high densities using pion, kaon, eta and photon productions in heavy-ion collisions}
\author{Zhi-Gang Xiao\inst{1,2} \and Gao-Chan
Yong\inst{3} \and Lie-Wen Chen\inst{4}  \and Bao-An Li\inst{5} \and
Ming Zhang\inst{6} \and Guo-Qing Xiao \inst{3} \and Nu Xu \inst{7} }

\institute{Department of Physics, Tsinghua University, Beijing
100084, China \and Collaborative Innovation Center of Quantum
Matter, Beijing, China \and Institute of Modern Physics, Chinese
Academy of Sciences, Lanzhou 730000, China \and Department of
Physics and Astronomy, Shanghai Jiao Tong University, Shanghai
200240, China \and Department of Physics and Astronomy, Texas A\&M
University-Commerce, Commerce, Texas 75429-3011, USA \and National
Institute of Metrology, Beijing 100013, China \and Institute of
Particle Physics, Central China Normal University, Wuhan 430079,
China }

\date{Received: date / Revised version: date}

\abstract{The high-density behavior of nuclear symmetry energy is
among the most uncertain properties of dense neutron-rich matter.
Its accurate determination has significant ramifications in
understanding not only the reaction dynamics of heavy-ion reactions
especially those induced by radioactive beams but also many
interesting phenomena in astrophysics, such as the explosion
mechanism of supernova  and the properties of neutron stars. The
heavy-ion physics community has devoted much effort during the last
few years to constrain the high-density symmetry using various
probes. In particular, the \rpi~ ratio has been most extensively
studied both theoretically and experimentally. All models have
consistently predicted qualitatively that the \rpi~ ratio is a
sensitive probe of the high-density symmetry energy especially with
beam energies near the pion production threshold. However, the
predicted values of the \rpi~ ratio are still quite model dependent
mostly because of the complexity of modeling pion production and
reabsorption dynamics in heavy-ion collisions, leading to currently
still controversial conclusions regarding the high-density behavior
of nuclear symmetry energy from comparing various model calculations
with available experimental data. As more \rpi~ data become
available and a deeper understanding about the pion dynamics in
heavy-ion reactions is obtained, more penetrating probes, such as
the \rki~ ratio, $\eta$ meson and high energy photons are also being
investigated or planned at several facilities. Here, we review some
of our recent contributions to the community effort of constraining
the high-density behavior of nuclear symmetry energy in heavy-ion
collisions. In addition, the status of some worldwide experiments
for studying the high-density symmetry energy, including the
 HIRFL-CSR external target experiment (CEE)  are briefly introduced.
 \PACS{
      {symmetry energy},
      {supra-saturation density}
     }
}

\maketitle

\section{Introduction}
\label{intro} Over a very wide beam energy range from tens of \amev~
to several TeV/u, one of the main purposes of heavy-ion collisions
is to study the bulk properties of strong interaction matter and
understand the QCD phase diagram. In the hardonic phase, the
symmetry energy term \esym in the equation of state (EOS) of
isospin-asymmetric nuclear matter (with unequal fractions of
neutrons and protons) is presently subject to the largest
uncertainty among all properties of nucleonic matter. The density
dependence of nuclear symmetry energy is an essential input for
calculating the properties of neutron stars. Moreover, it is also
important for understanding nuclear structure and reactions,
particularly those involving exotic nuclei \cite{lba08,lba11}. While
some significant progress has been made over the last few years in
constraining the \esym~ at sub-saturation densities using both
terrestrial nuclear laboratory data and astrophysical observations
\cite{tsa12,Lat12,ChenLW12,LiBA12}, still very little is known about
the \esym~ at supra-saturation densities.

Thanks to the collaborative effort of the community during the last
decade, several promising probes of the symmetry energy at
supra-saturation densities have been identified, such as the \rpi
\cite{lba05,qfjpg05} and \rki \cite{qfjpg05,fer06} ratios, the
neutron-proton differential transverse \cite{lba02} and elliptic
flows \cite{rus11}, the neutron/proton ratio of squeezed-out
nucleons~\cite{yong07}, triton over $^3$He ratio
\cite{ChenLW03,yong09}, $\Sigma^{-}/\Sigma^{+}$ ratio~\cite{LiQF05},
$\eta$ meson \cite{y2013} and high energy photon \cite{yong08,myg12}
productions. However, presently our conclusions regarding the
high-density behavior of nuclear symmetry energy are still
inconclusive, and in some cases controversial. For example, the FOPI
collaboration at GSI, Darmstadt published a few years ago a complete
set of pion data in heavy-ion collisions \cite{fopi07}. From
analyzing the charged pion ratio within the IBUU04 transport model
\cite{IBUU04}, a circumstantial evidence for a super-soft symmetry
energy was reported \cite{xzg09}. However, an analysis of the same
data using a version of the quantum molecular dynamics model gave an
opposite conclusion \cite{feng10}. More recent studies using the
Boltzmann-Langiven approach \cite{xie13} made a similar conclusion
as in ref. \cite{xzg09}. This situation has stimulated some
interesting debate and certainly calls for more studies, such as,
the in-medium effects of $\Delta$ and pion production
\cite{xuj10,yong10,xu13}, both experimentally and theoretically. It
is interesting to note that many theoretical studies  have been
performed to better understand the causes of the uncertain symmetry
energy at supra-saturation densities. In particular, effects of the
three-body force \cite{xuc10,wzuo}, isospin dependence of
short-range nucleon-nucleon correlation and tensor force
\cite{xuc13,Rho}, high order isospin asymmetry effect
\cite{clw09,Che11b} and exchange of $\rho-\omega$ mesons
\cite{jwz09}, deserve special attention. In this article, we
highlight a few key points reported in several of our recent studies
about probing the high-density symmetry energy using pion, kaon, eta
and high energy photon production in heavy-ion collisions. Several
newly planned experiments are also discussed.

The paper is arranged as following. Section 2 explains why heavy ion
collisions in terrestrial laboratory are useful for determining
\esym. Section 3 discusses the pion, kaon, $\eta$ and photon
productions and their relevance in constraining \esym~ at
supra-saturation densities. Section 4 introduces briefly the future
experiments planned worldwide and Section 5 is the summary.

\section{Symmetry energy and the isospin asymmetry of dense matter}
\label{ifractionation} 
\begin{figure}
\resizebox{0.5\textwidth}{!}{
  \includegraphics{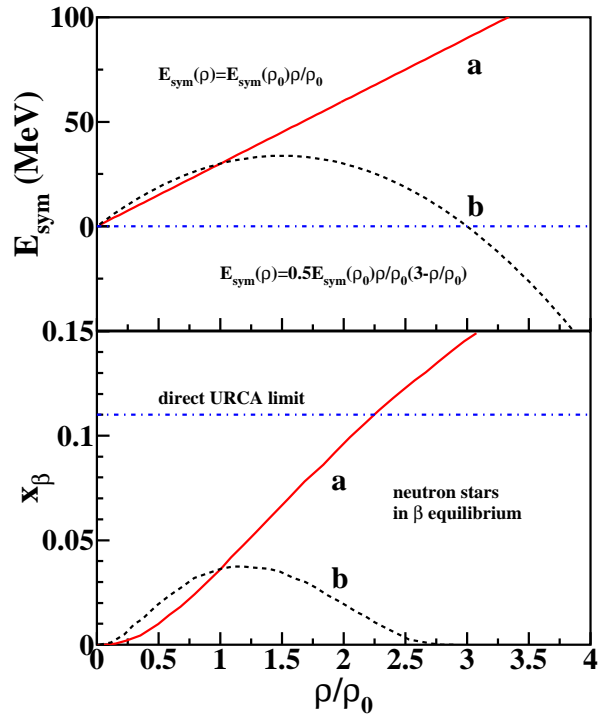}
} \caption{(color online) Upper window: Two representatives of the
nuclear symmetry energy as a function of density. Lower window: the
corresponding proton fractions in neutron stars at $\beta$
equilibrium. Taken from Refs. \cite{lba02,li02npa}.} \label{Li1}
\end{figure}

Why are the heavy-ion collisions useful for
probing the symmetry energy at supra-saturation densities? It is
necessary to first answer this question before we discuss potential
probes of the high-density symmetry energy in heavy-ion collisions.
The answer lies in the fact that the isospin asymmetry of dense
matter formed in heavy-ion collisions are intimately related to the
density dependence of nuclear symmetry energy. As an example, shown
in the upper window of Fig.\ \ref{Li1} are two possible functions of
the symmetry energy \esym. As discussed earlier in Refs.
\cite{lba02,li02npa}, the considered two forms of the symmetry
energy have the same value of $E_{\rm sym}(\rho_0)=30$ MeV at the
normal nuclear matter density $\rho_0$ and are very close to each
other at lower densities. At high densities they have completely
different trends reflecting the diverging predictions of nuclear
many-body theories. In the simplest model of neutron stars
consisting of $npe$ matter at $\beta$ equilibrium, the proton
fraction $x_{\beta}$ entirely determined by the \esym~ is shown in
the lower window of Fig.\ \ref{Li1}. With the $E^{b}_{\rm
sym}(\rho)$, the $x_{\beta}$ is zero for $\rho/\rho_0\geq 3$, while
with the $E^{a}_{\rm sym}(\rho)$, the neutron star becomes so
proton-rich that the fast cooling can happen at densities higher
than about $2.3\rho_0$.

\begin{figure}
\resizebox{0.5\textwidth}{!}{
  \includegraphics{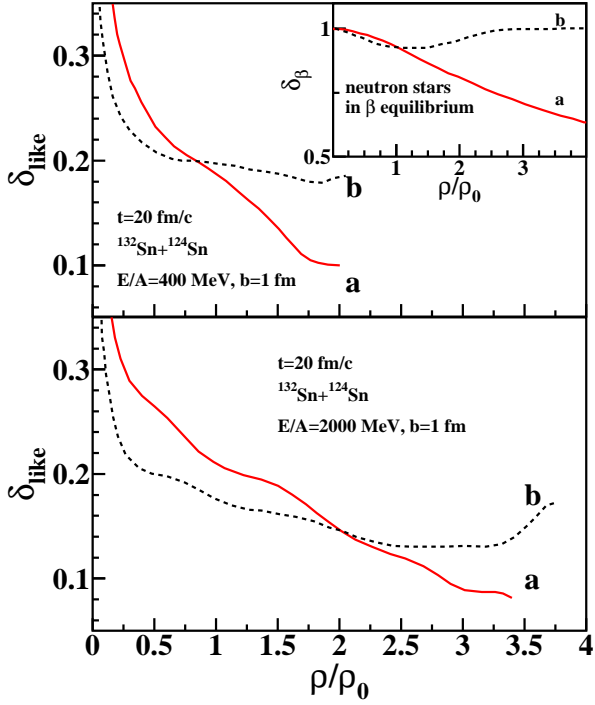}
} \caption{ (color online) Upper window: the isospin
asymmetry-density correlations at t=20 fm/c and $E_{\rm beam}=400$
\amev~ in the central $^{132}$Sn+$^{124}$Sn reaction with the
nuclear symmetry energy $E^a_{\rm sym}$ and $E^b_{\rm sym}$,
respectively. Lower window: the same correlation as in the upper
window but at 10 fm/c and $E_{\rm beam}=2$ \agev. The corresponding
correlation in neutron stars is shown in the insert. Taken from
Refs. \cite{lba02,li02npa}.} \label{Li8}
\end{figure}

Similar to the isospin fractionation in liquid-gas phase transition
in heavy-ion collisions at intermediate energies
\cite{Serot,LiKo,hushan00}, the different high-density behaviors of
the symmetry energy also lead to very different isospin asymmetry in
heavy-ion collisions. As examples, shown in Fig.\ \ref{Li8} are the
correlations between the baryon density and the isospin asymmetry
$\delta_{like}$ over the entire reaction volume at the time of about
the maximum compression in the central $^{132}$Sn+$^{124}$Sn
reactions with beam energy of 400 (upper window) and 2000 (bottom
window) \amev, respectively. How the initial $\rho-\delta$
correlation evolves in a heavy ion collision depends sensitively on
the high-density behavior of nuclear symmetry energy. With the
$E^a_{\rm sym}(\rho)$ the continuous neutron distillation from
higher density regions to lower ones persists at all energies. For
instance,the isospin asymmetry of the high-density region around
$\rho=2\rho_0$ with $E^b_{\rm sym}(\rho)$ is about twice  that with
the $E^a_{\rm sym}(\rho)$ at 400 \amev.
It is very interesting to note the astonishing similarity in the
resultant $\delta-\rho$ correlations for neutron stars and heavy-ion
collisions. In both cases, the symmetry energy $E^b_{\rm sym}(\rho)$
makes the high-density nuclear matter more neutron-rich than the
$E^a_{\rm sym}(\rho)$ and the effect grows with the increasing
density \cite{lba02,li02npa}.

\section{Particle production in heavy-ion collisions as potential probes of the symmetry energy at supra-saturation densities}
\label{probes} In the following we review a few recent studies about
using pion, kaon and $\eta$ mesons  as well as high energy photons
as potential probes of the high-density symmetry energy.
\subsection{The \rpi~ ratio as a probe of nuclear symmetry energy}
\label{pion}

\begin{figure}

\resizebox{0.5\textwidth}{!}{
  \includegraphics{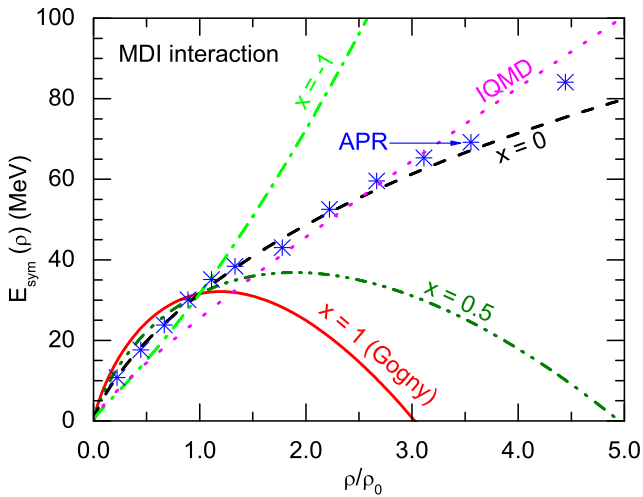}
} \caption{(Color Online) The symmetry energy as a function of
densities derived from the modified Gogny momentum dependent
interaction. The well-known APR prediction \cite{apr98} (blue
asterisks) and the parametrization used in IQMD model \cite{har98}
(pink dotted) are also plotted for comparison. Taken from ref.
\cite{xzg09}.} \label{esym_curve}
\end{figure}

As discussed in detail  in the literature, see, e.g., Refs.
\cite{lba02,li02npa,lba03,Fer05,xie13,cozma11}, the \rpi~ ratio in
heavy-ion collisions may be used as a sensitive probe of the
high-density behavior of nuclear symmetry energy. In fact, it was
realized a long time ago that the \rpi~ ratio depends strongly on
the isospin asymmetry of the reaction system, see, e.g., Refs~
\cite{ben79,nag81,har85,sto86}. This can be understood qualitatively
from both the $\Delta$ resonance model and the statistical model for
pion production.  Within the $\Delta$ resonance model for pion
production from first-chance independent nucleon-nucleon
collisions\cite{sto86}, the primordial \rpi~ ratio is
$(5N^2+NZ)/(5Z^2+NZ)\approx (N/Z)^2$. It is thus a direct measure of
the isospin asymmetry $(N/Z)_{\rm dense}$ of the dense matter in the
participant region of heavy-ion collisions. As we have discussed
earlier, the $(N/Z)_{\rm dense}$ is uniquely determined by the
high-density behavior of the nuclear symmetry energy. Therefore, the
\rpi~ ratio can be used to probe sensitively the high-density
behavior of nuclear symmetry energy. On the other hand, within the
statistical model for pion production\cite{nature}, the \rpi~ ratio
is proportional to ${\rm exp}\left[(\mu_n-\mu_p)/T\right]$, where
$T$ is the temperature, $\mu_n$ and $\mu_p$ are the chemical
potentials of neutrons and protons, respectively. At modestly high
temperatures ($T\geq 4$ MeV), the difference in the neutron and
proton chemical potentials can be written as \cite{thermal}
\begin{eqnarray}
\mu_n-\mu_p&=&V^n_{asy}-V^p_{asy}-V_{\rm Coulomb}\\ \nonumber
&+&T\left[{\rm
ln}\frac{\rho_n}{\rho_p}+\sum_m\frac{m+1}{m}b_m(\frac{\lambda_T^3}{2})^m(\rho^m_n-\rho^m_p)\right],
\end{eqnarray}
where $V_{\rm Coulomb}$ is the Coulomb potential for protons,
$\lambda_T$ is the thermal wavelength of a nucleon and $b'_m$s are
the inversion coefficients of the Fermi distribution
function\cite{thermal}. The difference in neutron and proton
 potentials $V^n_{asy}-V^p_{asy}=2v_{asy}(\rho)\delta$,
where the symmetry potential $v_{asy}(\rho)$ is directly related to
the symmetry energy~\cite{Xuli10b,xulinpa,Che12a}. It is seen that
the difference $\mu_n-\mu_p$ relates directly to the isospin
asymmetry $\rho_n/\rho_p$ or $\rho_n-\rho_p$. Thus within the
statistical model, the \rpi~ ratio is also sensitive to the
$(N/Z)_{\rm dense}$. Moreover, the value of $\pi^-/\pi^+$ ratio is
affected by the competition of the symmetry and Coulomb potentials
which all depend on the isospin asymmetry of the reaction system.
While the expectations based on these two idealized models provide a
useful guide, advanced transport model calculations are necessary.
Indeed, extensive studies have been done using various versions of
 transport models for haevy-ion collisions. One
such model is the isospin- and momentum- dependent
Boltzmann-Uehling-Uhlenbeck transport model (IBUU04). In this model,
the isospin-dependent in-medium nucleon-nucleon (NN) elastic
scattering cross sections are adopted, while the inelastic cross
sections are extracted from experimental free-space NN inelastic
collisions. For the single nucleon potential, the modified Gogny
momentum dependent interaction (MDI) is used. With this potential
the neutron effective mass is higher than the proton effective mass
in neutron-rich nuclear matter and the splitting between them
increases with both the density and isospin asymmetry of the medium.
By integrating over the momentum, the symmetry energy as a function
of density is then derived and plotted in Fig. \ref{esym_curve}. The
variable $x$ is introduced to mimic different forms of the \esym~
predicted by various many body theories without changing any
property of the symmetric nuclear matter and the symmetry energy at
normal density ~\cite{Che05a}. For comparison, the well-known
Akmal-Pandharipande-Ravenhall (APR) prediction \cite{apr98} (blue
asterisks) and the parametrization used in the Isospin-dependent
Quantum Molecular Dynamics (IQMD) model \cite{har98} (pink dotted)
are also presented.

\begin{figure}
\resizebox{0.52\textwidth}{!}{
  \includegraphics{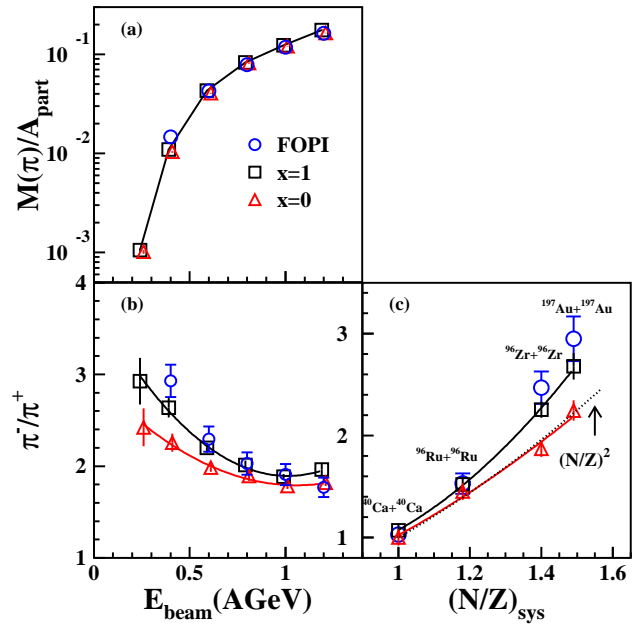}
} \caption{(Color Online)  The excitation function of the total
yield of charged pions (a) and of the yield ratio \rpi~ (b). (c) The
yield ratio \rpi~ as a function of the system $N/Z$. Taken from ref.
\cite{xzg09,xzg11}.} \label{pion_probe}
\end{figure}

Comparisons of the FOPI data and the IBUU04 calculations are shown
in Fig. \ref{pion_probe}. The window (a) displays the reduced
multiplicity of the charged pions $M_\pi/A_{\rm part}$ where $A_{\rm
part}$ denotes the number of participant nucleons. It is seen that
the total pion yield exhibits insignificant dependence on the
symmetry energy parameter $x$. However, if we inspect the pion yield
ratios \rpi~ as a function of beam energy (b) and of system isospin
composition (c), it is interesting to notice that the FOPI data
favors clearly the calculation with x = 1 corresponding to a very
soft symmetry energy. The calculation with $x=0$ is consistent with
the IQMD calculations but deviates considerably from the data
\cite{xzg09}. As discussed above, an enhanced pion ratio \rpi~ means
that more neutrons are resided in the high-density region to produce
more $\pi^-$ according to the simple isobar model. To achieve a
neutron rich dense region, a soft symmetry energy is required. It is
worth noting that the super-soft symmetry energy can not support
massive neutron stars with the normal EOS. This has triggered the
interesting application of non-Newtonian gravity in neutron stars
\cite{wen09}. On the other hand we would also note that the symmetry
energy sensitivity of the \rpi~ ratio  only shows up below 600
\amev~ corresponding to a nuclear matter density approximately lower
than $2\rho_0$.

It should be pointed out that model dependent conclusions have been
reported based on the same data set from FOPI collaboration. Using
an isospin dependent quantum molecular dynamics (IQMD) model, Feng
et al. have found that in order to reproduce the data a rather stiff
symmetry energy is necessary \cite{feng10}. In another analysis done
recently by Xie et al. using the Boltzmann-Langevin equations, a
very soft symmetry energy similar with the curve represented by
$x=1$ in IBUU04 analysis is favored \cite{xie13}.

\begin{figure}
\resizebox{0.5\textwidth}{!}{
  \includegraphics{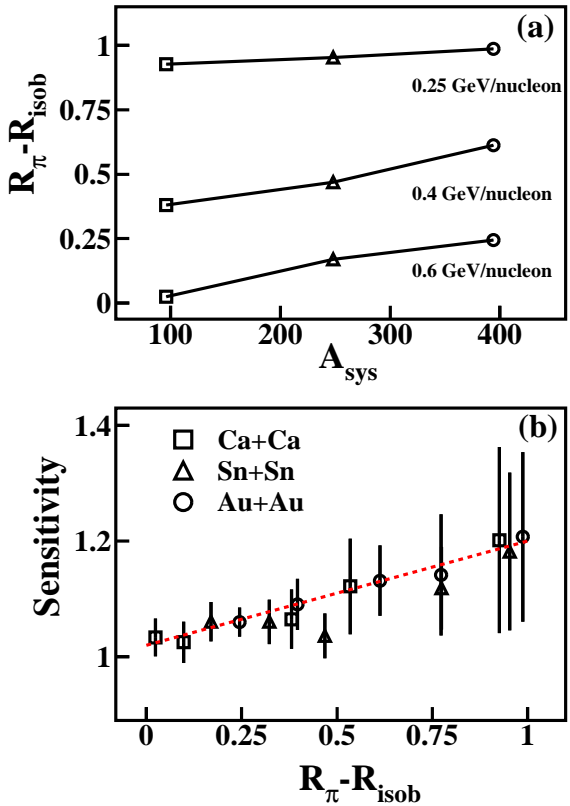}
} \caption{(Color online) (a) the system size and beam energy
dependence of the degree of isospin fractionation denoted by
$R_{\pi}-R_{\rm isob}$ (see text). (b) correlation between the
sensitivity of the \rpi~ ratio to the \esym~ and the degree of
isospin fractionation. The dashed line is for guiding the eyes.
Taken from Ref. \cite{zm09} } \label{fractionation}
\end{figure}

To arrive at a convincing constraint of the symmetry energy at
supra-saturation densities from the pionic probe, more theoretical
and experimental studies are required. From experimental point of
view, it is of significance to optimize the conditions such as the
reaction system and the beam energies to achieve the maximum
sensitivity that the probes depend on the symmetry energy.
Qualitatively thinking, the symmetry energy takes effect if the
isospin fractionation mechanism is at work. The deeper the degree of
isospin fractionation, the more sensitive the dependence of the
probe on the symmetry energy probe. To check whether the isospin
fractionation is relevant to the space-time volume of the colliding
system, the initial idea is to find a quantity that can be used to
measure the degree of isospin fractionation. While the isobar model
predicts the \rpi~ ratio based on the static $N/Z$ composition of
the system with $R_{\rm isob}=\frac{5N^2 + NZ}{5Z^2+NZ}$
\cite{sto86}, the transport model gives the \rpi~ ratio $R_{\pi}$
based on the dynamic simulation of the compression (leading to high
density) and the expansion (leading to low density) of the whole
stage. Thus, to the first order of approximation, we adopt their
difference $R_{\pi}-R_{\rm isob}$ to represent the degree of isospin
fractionation and plot it in  Fig. \ref{fractionation} (a) as a
function of the mass $A_{\rm sys}$ for three systems with same $N/Z$
ratio, \caca, \snsn~ and \auau, at beam energies of 0.25, 0.4 and
0.6 \agev, respectively. The symmetry energy is parameterized with
$x=1$ in the transport model calculation. It is seen that
$R_{\pi}-R_{\rm isob}$ exhibits a positive correlation to the system
size at each beam energy, indicating a dependence of the isospin
fractionation on the space-time volume of the colliding system. At
lower beam energy, as the maximum density achieved is lower while
the reaction time is longer, the net effect of the isovector
potential on the \rpi~ ratio is larger. At higher energies, besides
the same trend, the $R_{\pi}-R_{\rm isob}$ exhibit a more rapid
increasing with $A_{\rm sys}$ due to probably a larger isovector
density gradients difference reached in the reactions. The
sensitivity of the \rpi ratio on the \esym, defined here by the
double ratio of the \rpi~ obtained with the \esym~ of $x=1$ over
that with $x=0$, is plotted in Fig. \ref{fractionation}(b) as a
function of the $R_{\pi}-R_{\rm isob}$. Interestingly it is shown
that the sensitivity  increases with the $R_{\pi}-R_{\rm isob}$
despite of the larger uncertainty experienced at lower beam energies
due to the low statistics of pions in the simulation.  Thus it is
suggested that the \rpi~ ratio in heavy systems near the pion
production threshold is preferred for probing the high densities
behavior of the \esym \cite{zm09}.

\begin{figure}
\resizebox{0.5\textwidth}{!}{
  \includegraphics{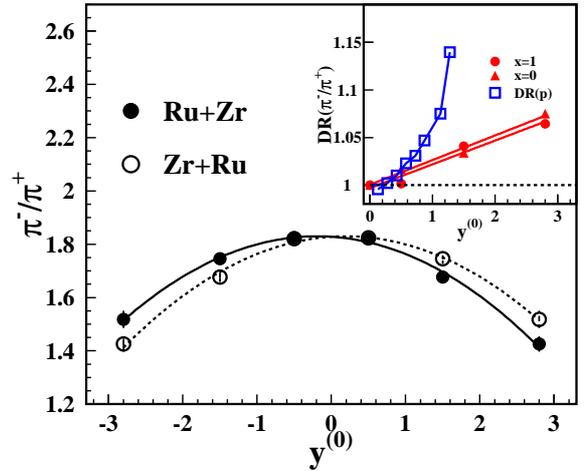}
} \caption{ (Color online) $\pi^-/\pi^+$ ratio as a function of
reduced rapidity $|y^{(0)}|$ in IBUU04 calculation with  $x=1$  in
$^{96}$Ru+$^{96}$Zr (solid, calculated) and $^{96}$Zr+$^{96}$Ru
(open, reflected by the solid symbols  with respect to the
midrapidity). Taken from \cite{zm10}. The insert shows the double
ratio $DR(\pi^-/\pi^+)$  in the two systems with $x$= 0 and 1 in
comparison with the experimental ratio of proton $R(P)$
\cite{hong02}(open square) } \label{tracer}
\end{figure}

It is of significance to clarify that the effect of the symmetry
energy on the  \rpi~ ratio originates spatially  from the high
density region. M. Zhang et al. compares the effect of changing the
\esym~ parameter in high and low densities separately, suggesting
that the high-density symmetry energy contributes about 80\% of the
total effect \cite{zm09}. This conclusion is consistent with an
earlier calculation using URQMD transport model \cite{qfjpg05}. The
isospin tracer with \rpi~ as a function of rapidity may provide an
indirect experimental test. According to transport simulation, the
$\Delta$ resonances are mainly produced in high-density region,
which is formed by the participants at earlier collision stage.
Thus, in a simple picture, the pions are less influenced by the
Corona effect than nucleons, for which the contribution from the
surface is not negligible \cite{zm10}. To see this effect, we study
the pion ratio \rpi~ in Zr+Ru and Ru+Zr collisions as a function of
the reduced rapidity, as shown in Fig. \ref{tracer}. It is seen that
a difference between the forward rapidity and backward rapidity is
evident, suggesting the isospin equilibrium is not achieved. More
interestingly one can see from the insert of Fig. \ref{tracer}  that
the double ratio of charged pions DR(\rpi) in Ru+Zr and Zr+Ru
systems shows approximately linear increase with rapidity while the
ratio of proton in these two systems exhibits a more rapid increase
with rapidity. The different behavior between the pions and the
protons suggests that the surface effect strongly experienced by the
later is less pronounced for the pions. If this effect is confirmed
by the experiment, it suggests that the primordial pions are mainly
produced in the high-density region, which is in agreement with
earlier transport model calculations \cite{zm09}.

\subsection{The \rki ratio as a probe to nuclear symmetry energy}
\label{kaon}

Compared with pions, kaons are also produced in the dense phase in
heavy-ion collisions but experience less final state interactions
and thus provide a useful probe of the EOS of nuclear matter at high
densities \cite{Aic85}. Earlier, the kaon yield has been used
successfully in constraining the incompressibility of isospin
symmetric nuclear matter. For instance, transport model analyses of
the Kaos data suggest that the EOS of symmetric matter is soft
\cite{kaos01,hat06,fuchs01}. For the isospin effect, Q. Li et al has
noticed in $^{132}$Sn+$^{132}$Sn at 1.5 \agev~ that the \rki~ ratio
depends on the symmetry potential \cite{qfjpg05}. Later in
\cite{fer06}, an excitation function of \rki in Au+Au collisions
with different symmetry potential calculated and confirmed the
dependence of the \rki ~ ratio on the isovector potential. It has
been realized that the isovector part of the in-medium interaction
affects the kaon production via two mechanisms: (i) a symmetry
potential effect, i.e., a larger neutron repulsion in n-rich
systems, and (ii) a threshold effect, due to the change in the
self-energies of the particles involved in inelastic processes. Thus
the \rki~ ratio was proposed as another, probably better, probe of
the high-density behavior of nuclear symmetry energy \cite{fer06}.
Similar to the \rpi~ observable, the sensitivity of the \rki~ to the
variation of symmetry energy is more pronounced at beam energies
near the kaon production threshold \cite{qfjpg05,fer06}.

\begin{figure}
\resizebox{0.5\textwidth}{!}{
  \includegraphics{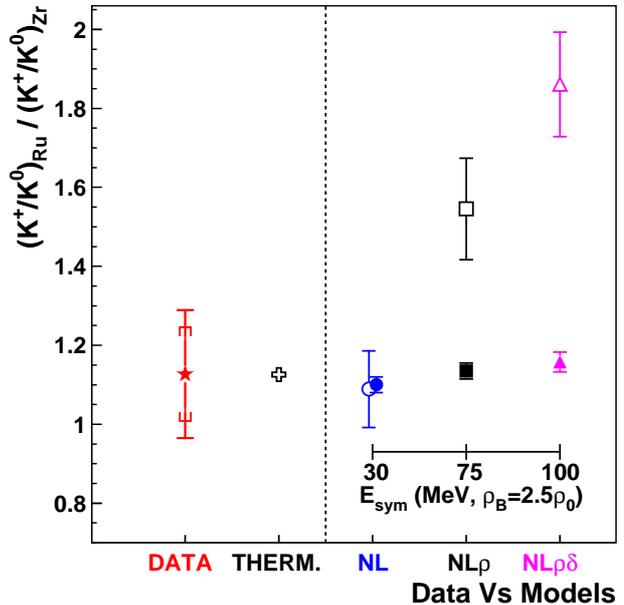}
} \caption{(Color Online)  The double ratio of \rki measured in
Ru+Ru and Zr+Zr systems (red star) in comparison with thermal model
calculations ( open cross in left panel) and  relativistic mean
field transport model (right panel) for infinite nuclear system
(open) and realistic collisions (solid), taken from
Ref.~\cite{lop07}. } \label{kaon_probe}
\end{figure}

The first attempt to extract the stiffness of symmetry energy from
\rki~ observable was done by the FOPI collaboration \cite{lop07}. To
reduce the systematic uncertainties, the double ratio of \rki~ in
Zr+Zr and Ru+Ru systems with same mass but characterized by
different $N/Z$ composition at 1.528 \agev~ is used. The
experimental result is compared with thermal model calculation and a
relativistic mean field transport model using two different
collision scenarios (the infinite nuclear matter and the realistic
finite nuclear collision) and under different assumptions on the
stiffness of the symmetry energy, as presented in Fig.
\ref{kaon_probe}. Thermal model calculation reproduces the data. For
the transport model calculations, the sensitivity of the double
ratio \rki~  on the stiffness of symmetry energy differs in the two
scenarios. In the former case, it is evident that the double ratio
\rki~  increases rapidly if one goes from $NL$ (symmetry energy only
containing kinetic contribution) via $NL\rho$ (including the
isovector-vector field $\rho$) to $NL\rho\delta$ (including both
$\rho$ and $\delta$ effective field), or equivalently with
increasing the stiffness of \esym~ from 50 to 100 MeV at
2.5$\rho_0$. While in the realistic nuclear collisions of Ru/Zr
nuclei,  the sensitivity is less pronounced and the double ratio of
\rki~ differs by about 5\% between $NL$ and $NL\rho\delta$
parameterizations. This reduced sensitivity is mainly attributed to
two dynamical effects that tend to weaken the contribution of the
isovector interaction, i.e., the fast neutron emission and the
transformation of neutron into proton in inelastic channels. It
indicates that to obtain experimentally accessible sensitivity, one
shall use the lower beam energy in heavier system with larger $N/Z$
asymmetry.  On the experimental side, however, measuring a sample of
$K^0$ and $K^+$ with sufficiently high statistics and high precision
at subthreshold energies will be a considerable challenge and calls
for great effort in future experiment.

\subsection{\textbf{$\eta$} production and effects of nuclear
symmetry energy} This subsection is a brief summary of the work
originally published in ref. \cite{y2013}. Given the inconsisteny of
the high density \esym~ among various transport model analyses,
searching for more sensitive new probes of \esym~ at high-densities
is still ongoing. Compared to pion or kaon, massive mesons generally
probe the \esym~ at higher densities if the effects of the final
state interactions do not smear the signal. The $\eta$ meson, as the
massive member in the nonet
of pseudoscalar Goldstone mesons with mass
547.853 MeV/c$^2$ \cite{eta08}, is a preferred probe to \esym~ for
further good reasons: (1) The $\eta$ meson experiences weaker final
state interactions compared to pion due to its hidden strangeness
(the $s\bar{s}$ component). And because the net strangeness content
is zero in $\eta$ mesons, they can be produced without another
strange partner in the final state and thus require less energies.
In addition, it has significant photon and dilepton decay branch
ratio, which provides a clean electromagnetic probe to \esym. (2)
The $\eta$ meson is sensitive to the number of p-n collisions while
the \rpi~ ratio is determined by the ratio of n-n and p-p colliding
pairs, $\eta$ and the \rpi~ ratio provide complementary information
about the symmetry energy. (3) The $\eta$ meson is produced from the
proton-neutron collision for which the
 relative momentum is determined mainly by the
gradient of the isovector potential. However, we shall keep in mind
that the elementary $\eta$ production cross sections in
baryon-baryon and meson-baryon scattering still suffer from larger
uncertainties although they are gradually better known as more data
and calculations become available \cite{wolf93,Tho07}. Moreover, the
detection of the photonic decay or Dalitz decay (with dilepton) of
$\eta$ is also challenging near the production threshold.

\begin{figure}
\resizebox{0.5\textwidth}{!}{
  \includegraphics{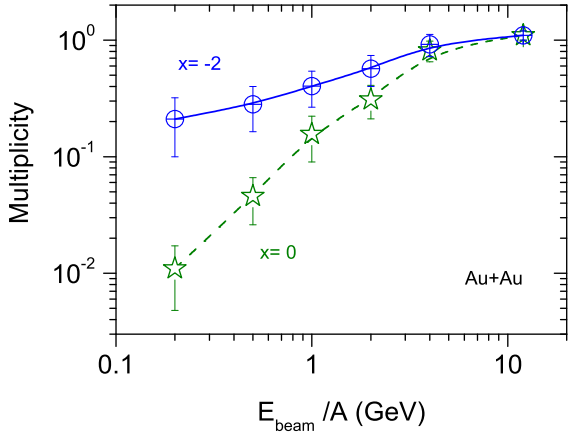}
} \caption{(Color online) Effects of the symmetry energy on the
multiplicity of inclusive $\eta$ production as a function of
incident beam energy in Au+Au reactions using a relativistic
transport model ART1.0. Figure is taken from Ref. \cite{y2013}. }
\label{exc}
\end{figure}

The sub-threshold $\eta$ production requires multiple
nucleon-nucleon scattering over sufficiently long time \cite{dep89}
and hence carries mainly the information of the equilibrium phase.
Fig.~\ref{exc} presents the $\eta$ multiplicity as a function of
beam energy in inclusive Au+Au reactions with both soft ($x= 0$) and
stiff ($x= -2$) symmetry energies using a relativistic transport
model ART1.0~\cite{LiBA95}. One can clearly see that the $\eta$
multiplicity decreases rapidly with decreasing the beam energy
especially for the soft symmetry energy but saturates at about 10
\agev~ with increasing the beam energy.  Over the whole energy range
under investigation, more (less) $\eta$ mesons are produced with the
stiff (soft) symmetry energy. Again due to the isospin fractionation
mechanism in the equilibrium phase, the symmetry energy and the
isospin asymmetry $\delta$ satisfies the equation $E_{\rm
sym}(\rho_{1})\delta_{1} = E_{\rm sym}(\rho_{2})\delta_{2}$ where
$\rho_{1}$ and $\rho_{2}$ represents two density conditions,
respectively \cite{shi00,li02npa}. Thus, with a higher symmetry
energy \esym~ the dense region is less isospin asymmetric and more
$np~(pn)$ collisions can occur to produce more $\eta$ mesons.
Interestingly, the effect of the nuclear symmetry energy becomes
stronger and stronger with decreasing beam energy and the
sensitivity of the $\eta$ production on \esym~ is more pronounced in
comparison to the probe of \rpi~ ratio in the same energy region.

\begin{figure}
\resizebox{0.5\textwidth}{!}{
  \includegraphics{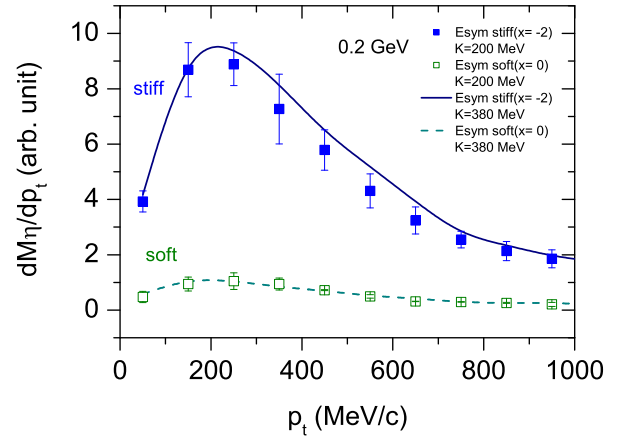}
}\caption{(Color online) Effects of the symmetry energy and the
incompressibility of symmetric nuclear matter on the transverse
momentum distributions of $\eta$ production in inclusive Au+Au
reaction at incident beam energy of 200 \amev~ using a relativistic
transport model ART1.0. Figure is taken from Ref. \cite{y2013}.}
\label{tran}
\end{figure}

Fig.~\ref{tran} further shows the transverse momentum distribution
of $\eta$ in Au+Au reactions at beam energy of 200 \amev~ with
different density-dependent symmetry energies and incompressibility
coefficients $K$. The ordinate $dM_{\eta}/dp_{t}$ represents the
differential $\eta$ multiplicity defined by $\frac{M_{\eta}(p_{t}
\rightarrow p_{t}+dp_{t})}{dp_{t}}$. By comparing the two curves it
is seen that the effect of \esym~ is most pronounced in the $p_t$
interval from about 200 to 300 MeV/c. The predictions with two
\esym~ parameters of $x=0$ and -2 differs by a factor of about 5 to
10, while the effect of changing the incompressibility of nuclear
matter from $K= 200$ MeV to 380 MeV is insignificant. This is
attributed to the fact that the p-n relative momentum is essentially
determined by the gradient of the isovector potential. For the
isoscalar part, the effects of the incompressibility  $K$ of the
symmetric nuclear matter on the production and the reabsorption of
$\eta$ nearly cancel out. It can be understood that a large $K$
causes less resonances (and hence less $\eta$) to be produced, on
the other hand a hard symmetric matter undergoes a fast expansion
causing less reabsorption. The effect of utilizing a momentum
dependent isoscalar potential is worthy of further investigation.

\subsection{High energy photon production and effects of nuclear
symmetry energy} \label{photon} This subsection is a brief summary
of the work originally published in ref. \cite{yong08}. The hadronic
probes of the nuclear symmetry energy  usually suffer from strong
final state interactions (FSI), although some special treatment,
such as using the squeezed-out neutron/proton ratio perpendicular to
the reaction plane \cite{yong07}, tends to reduce strong FSI effect.
To overcome this difficulty, the electromagnetic observables, such
as the photons or dileptons , are proposed as cleaner probes to the
nuclear symmetry energy \esym~, especially at supranormal densities.

During the last twenty years, hard photon emissions in heavy ion
reactions at various beam energies have been extensively studied
both theoretically and experimentally by several
groups\cite{bertsch88,nif90,cassrp}. It has been concluded that the
high energy $\gamma$ rays are mainly produced in the early stage of
the reaction and have very low probability to experience final state
interaction. Thus besides the symmetry energy effect, the emitting
hard photons can probe not only the reaction dynamics leading to the
formation of dense matter
\cite{bertsch86,ko85,cassing86,bau86,stev86}, but also the in-medium
NN (nucleon-nucleon) scattering cross sections \cite{yong11}.
Nevertheless, the major uncertainty of the probability $p_{\gamma}$
of the input elementary process $pn\rightarrow pn\gamma$ , which is
a major concern so far, is still rather model
dependent\cite{nif85,nak86,sch89,gan94,tim06}.

\begin{figure}
\resizebox{0.5\textwidth}{!}{
  \includegraphics{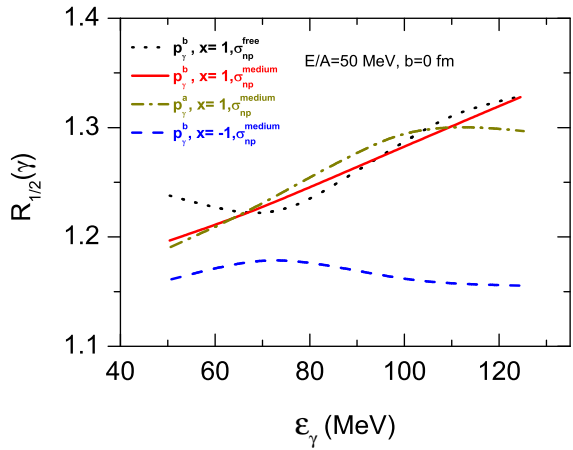}
}\caption{(Color online) The spectra ratio of hard photons in the
reactions of $^{132}Sn+^{124}Sn$ and $^{112}Sn+^{112}Sn$ reactions
at a beam energy of $50$ \amev~ using IBUU04 transport model with
interaction MDI($x=-1$) (very stiff symmetry energy) and MDI($x=1$)
(very soft symmetry energy), respectively. Taken from Ref.
\cite{yong08}.} \label{relat}
\end{figure}
Since a double ratio  of a certain observable in two reactions
usually reduce the systematic uncertainty and other ``unwanted
effects", as early discussed in Ref. \cite{yong08}, we have proposed
to measure the spectra ratio $R_{1/2}(\gamma)$ of hard photons from
two head-on reactions of $^{132}$Sn+$^{124}$Sn and
$^{112}$Sn+$^{112}$Sn, i.e.,
\begin{equation}
R_{1/2}(\gamma)\equiv\frac{\frac{dN}{d\varepsilon_{\gamma}}(^{132}\rm
Sn+^{124}\rm Sn)} {\frac{dN}{d\varepsilon_{\gamma}}(^{112} \rm
Sn+^{112}\rm Sn)}.
\end{equation}
By choosing the two systems with approximately equal mass but rather
different isospin compositions, we hope to see the isospin effect.
Shown in Fig.\ \ref{relat} is the $R_{1/2}(\gamma)$ calculated using
IBUU04 transport model with different symmetry energy stiffnesses,
i.e., MDI( $x=-1$) (very stiff) and MDI($x=1$) (very soft),
respectively. To show the effects arising from the two body
scattering, the process probability of the elementary process
$pn\rightarrow pn\gamma$ is taken in two forms, $p^a_{\gamma}$ and
$p^b_{\gamma}$ \cite{yong08} and the NN cross section is also varied
in the calculation with $\sigma_{\rm np}^{\rm free}$ being the cross
section in free space and $\sigma_{\rm np}^{\rm medium}$ being the
value modified due to medium effects, respectively. As expected, the
full calculations with the $p^a_{\gamma}$ and $p^b_{\gamma}$ and
different NN cross sections indeed lead to about the same
$R_{1/2}(\gamma)$ within statistical uncertainties since the
corresponding effects are canceled out by the  ratio method.
Comparing the results with ($x=1$) and MDI($x=-1$) both using the
$p^b_{\gamma}$, however, it is clearly seen that the
$R_{1/2}(\gamma)$ remains sensitive to the symmetry energy
especially for the very energetic photons. This verify numerically
the advantage of using the $R_{1/2}(\gamma)$ as a robust probe of
the symmetry energy. Despite of the significant effect, we would
also note here that the detection of hard photons is also
challenging and requires careful treatment since the $\gamma$ rays
from other processes are involved in heavy ion collisions in the
relevant energy range.

\section{Perspective} \label{exp}

In order to have insight into the nuclear equation of state at
supra-saturation densities using heavy-ion collisions, it is a
prerequisite that a compressed phase of nuclear matter, which is the
environment where the \esym~ takes effect, can be formed and last
for adequately long time. In the collisions, not only the symmetry
potential as a part of the mean field, but also many other factors
like the in-medium modification of the hadrons properties and the
two body collision cross sections come into play. In addition, the
compression, expansion and cooling processes are dynamically
correlated in heavy-ion collisions. Transport models providing
consistent description of the multi facets of the collisions are
required to bridge the nuclear equation of state and the
experimental observables in order to draw model- and observable-
independent conclusions on the \esym.

From experimental side, we have shown that heavy-ion collisions in
sub-\agev~ regime is beneficial for the experimental constraining of
symmetry energy in terrestrial laboratory. Despite of many
experimental challenges, some worldwide radioactive beam facilities
and the experiments below 1 \agev~ bombarding energy and with larger
beam intensity are under constructions. For instance, the ASY-EOS
experiment at GSI in Germany is dedicated to measure the
differential neutron and proton flow in heavy ion collisions at
several hundreds \amev  \cite{nn2012,rus2013,traut2013}.  The
SAMURAI spectrometer has been constructed  at RIKEN in Japan
\cite{isobe11}, which was designed for kinematically complete
radioactive beam experiments, including the pion production
measurement in near threshold heavy ion collisions.  The
large-acceptance multipurpose spectrometer (LAMPS) is to be
constructed  on RAON in Korean \cite{hong13}. The high energy stage
of LAMPS combines a solenoid and a dipole with neutron detectors. It
will be able to measure charged pions, fragments and neutrons with
large coverage and hence provides further experimental data in
constraining the high density symmetry energy. Besides, the cooling
storage ring (HIRFL-CSR) in China, delivering heavy-ion beams up to
1 \agev \cite{xia02}, to be coupled with advanced detectors at the
external target experiment as called CEE (CSR External-target
Experiment), will also contribute to further studies of
\esym\cite{SQM08}. The central components of CEE are the tracking
detectors inside and downstream to a dipole with large acceptance at
forward and midrapidity in laboratory reference. For the forward
rapidity acceptance, it is expected to provide complementary data to
the existing devices \cite{zmcpc}. Meanwhile, further observables ,
for instance,  the balance energy of direct nucleon flow
\cite{gcc12}, the scattering angles of the nucleons on a heavy
target \cite{oul08}, are demonstrated sensitively probing the
symmetry energy at high densities. With the running of these
experiments and the cross check with more observables coming into
play, the next decade will hopefully witness more abundant data
leading to further constraints of the high-density behavior of
\esym.

\section{Summary}
\label{conclusion}

Some recent progress and remaining issues in constraining the
symmetry energy at supra-saturation densities in heavy-ion
collisions are reviewed. Large uncertainties of the behavior of the
symmetry energy beyond saturation density still exist both
theoretically and experimentally. Based on BUU and
Boltzmann-Langevin transport model analysis of the most recent
systematic pion data, circumstantial evidence suggesting a rather
soft symmetry energy is obtained. However, this conclusion is
different from those based on some other model analyses. Further
simulations suggest that large symmetry energy sensitivity of the
charged pion ratio can be expected in heavy-ion collisions below 0.5
\agev, supporting further studies at several new facilities. After
being proposed as a promising probe, Kaonic ratio \rki~ has been
measured in Ru/Zr collisions at 1.58 \agev~ but does not lead to a
convincing constraint of \esym~ because the symmetry energy effect
can not be discriminated within the current experimental
uncertainty. With the discrepancy of the high-density behavior of
the symmetry energy remaining unresolved, more sensitive probes of
relevance are always useful. $\eta$ and photon production are also
potential tools to constrain the symmetry energy. Further
experimental studies and theoretic work are required.

\section{Acknowledgement}

This work is supported by the CUSTIPEN (China-U.S. Theory Institute
for Physics with Exotic Nuclei) under DOE grant number
DE-FG02-13ER42025, the US National Science Foundation under Grant
No.  PHY-1068022,  the National Aeronautics and Space Administration
under grant NNX11AC41G issued through the Science Mission
Directorate,  the Natural Science Foundation of China under Grant
No. 11079025, 11135011,  11275125, 11375239, 11375094 and U1332207
Tsinghua University Initiative Scientific Research Program, the
Shanghai Rising-Star Program under grant number 11QH1401100, the
``Shu Guang" project supported by Shanghai Municipal Education
Commission and Shanghai Education Development Foundation, the
Program for Professor of Special Appointment (Eastern Scholar) at
Shanghai Institutions of Higher Learning, and the Science and
Technology Commission of Shanghai Municipality (11DZ2260700).


\end{document}